\documentstyle[12pt]{article}
\begin{document}
\topmargin -0.2cm \oddsidemargin -0.2cm \evensidemargin -1cm
\textheight 22cm \textwidth 12cm

\title{Creation of neutron spinless pairs in a superfluid liquid 
$^4$He and a neutron gas mixture.}  
\author{Minasyan V.N. \\
Yerevan, Armenia}

\date{\today}

\maketitle

\begin{abstract} 
First, the creation of a free neutron spinless pairs is predicted 
in a superfluid liquid $^4$He and a neutron gas mixture. For solving 
the given problem, it is presented an exact solution to 
the model of dilute Bose gas as an extension of the Bogoliubov model, at quantity of the condensate fraction varying in the state $0\leq\frac{N_0}{N}\leq 1$, which in turn might be useful for a description of the superfluid liquid $^4$He.  Due to an application of presented new model of dilute Bose gas, we prove that an appearance of atoms in the condensate is a suppressor for the collective modes as well as a creator for single-particle excitations On other hand, it is shown that the terms of the interaction between the Bogoliubov excitations (Bogoliubov phonon-roton modes) and the density modes of the neutron meditate an attractive interaction via the neutron modes, which in turn leads to a bound state on a spinless neutron pair. 
\end{abstract} 

PACS $67.40.-w$

\vspace{100mm}

\vspace{5mm} 
 
{\bf 1. Introduction.} 
 
\vspace{5mm} 

The motivation for our theoretical study of the very low-temperature properties of the dilute hard sphere Bose gas is an attempt at a microscopic understanding of superfluidity in helium $^4$He. We proceed by discussing some experimental and theoretical investigations. 

The connection between the ideal Bose gas and superfluidity in helium was first made by London [1] in 1938. The ideal Bose gas undergoes a phase transition at sufficiently low temperatures to a condition in which the zero-momentum quantum state is occupied by a finite fraction of the atoms. This momentum-condensed phase was postulated by London to represent the superfluid component of liquid $^4$He. With this hypothesis, the beginnings of a two- fluid hydrodynamic model of superfluids was developed by Landau [2] where he predicted the notation of a collective excitations so- called phonons and rotons. 

The purely microscopic theory with mostly utilizes-technique, was first described by Bogoliubov [3] within the model of weakly non-ideal Bose-gas, with the inter-particle S- wave scattering. Based on the application of the presence of a macroscopic number of condensate atoms $N_0\approx N$ (where $N$ is the total number of atoms), Bogoliubov has dropped a density operator term which describes the fluctuation of atoms above the zero momentum level, the Bogoliubov obtained the dispersion curve for single particle Bogoliubov excitations (Bogoliubov phonon-roton modes).   

The dispersion curve of an excitations excited in superfluid helium has been accurately measured a function from momentum [4]. Within this experiment, the position of a sharp peak inelastic neutron scattering intensity defines by energy of the   single particle excitations, and there is appearing a broad component in inelastic neutron scattering intensity, at higher momenta of atoms. For explanation of the appearance of a broad component in inelastic neutron scattering intensity, the authors of papers [5-7] proposed to consider the presence of collective modes in the superfluid liquid $^4$He, which are represented a density excitations. In this respect, the authors of this letter predicted that the collective modes represent as the density quasiparticles [8]. In these works, these density excitations and density quasiparticles are appeared due to remained density operator term for describing atoms above the condensate, which was neglected by Bogoliubov [3]. 

In this letter, we present a new model of a nonideal Bose gas for describing of the superfluid liquid helium. The given model is based on the application of the Penrose-Onsager definition of the Bose condensation [9] which is based on the condition for a condensed fraction of atoms  $\frac{N_0}{N}=const$. The later explains a broken of Bose-symmetry law for the atoms of the Bose gas in the condensate level, and gives a fully exact solution to the model of dilute non-ideal Bose gas, at quantity of the condensate fraction $0\leq\frac{N_0}{N}\leq 1$. In this context, the new model of a nonideal Bose gas, presented herein, leads to an absence of the collective modes because appearance of atoms in the condensate is the suppressor for collective modes as well as the creator for single-particle excitations. Therefore, we prove that the density excitations and the density quasiparticles, proposed by authors of letters [5-8], are unphysical.
 
The last time, the authors of letter [9] discovered an existence of scattering between atoms of the superfluid liquid helium, at lambda transition, which is confirmed by calculation of the dependence of the critical temperature on the interaction parameter as scattering length. On other hand, as it is indicated by authors of [4], there are two type excitations in superfluid helium, at lambda transition. These facts imply that it needs to revise the concept of determination condition for Bose-Einstein condensation in the superfluid liquid helium. Obviously, the peak inelastic neutron scattering intensity is connected with a registration of neutron modes by neutron-spectrometer, which in turn defines sort of excitations.

Further, we investigate a helium liquid-neutron gas mixture where exists the term of interaction between the Bogoliubov modes and the density modes of the neutrons, which due to application a canonical transformation of the Hamiltonian of system, the term of the interaction between the density of the Bogoliubov modes and the density of the neutron modes is removed by meditated an effective attractive interaction between the neutron modes, which in turn determines a bound state on neutron pair.
 
Therefore, in this letter, we present to determine a condition for Bose condensation, at lambda transition, by a registration of neutron-spectrometer as single neutron modes as well as neutron pair modes.

\vspace{5mm}
{\bf 2. New model of a dilute  Bose gas.}
\vspace{5mm}

For beginning, we present a new model of a dilute Bose gas for describing property of superfluid liquid helium. The given model considers a system of $N$ identical interacting atoms via S-wave scattering. These atoms, as spinless Bose-particles,  have a mass $m$ which are confined in a box of volume $V$. The main part of the Hamiltonian of such system is expressed in the second quantization form as: 

\begin{equation}
\hat{H}_a= \sum_{\vec{p}\not=0}\frac{p^2}{2m}
\hat{a}^{+}_{\vec{p}}\hat{a}_{\vec{p}}+
\frac{1}{2V}\sum_{\vec{p}\not=0}U_{\vec{p}}\hat{\varrho}_{\vec{p}}
\hat{\varrho}^{+}_{\vec{p}}
\end{equation}

Here  $\hat{a}^{+}_{\vec{p}}$ and $\hat{a}_{\vec{p}}$ are, respectively, the
"creation" and  "annihilation" operators of a free atoms with 
momentum $\vec{p}$;  $U_{\vec{p}}$ is the Fourier transform 
of a S-wave pseudopotential in the momentum space:

\begin{equation}
U_{\vec{p}} =\frac{4\pi d \hbar^2}{m }
\end{equation}

where $d$ is the scattering amplitude; and 
the Fourier component of the density operator presents as 

\begin{equation}
\hat{\varrho}_{\vec{p}}=\sum_{\vec{p}_1}
\hat{a}^{+}_{\vec{p}_1-\vec{p}}\hat{a}_{\vec{p}_1}
\end{equation}
  
According to the Bogoliubov's theory [3], it is a necessary to separate the atoms in the condensate from those atoms filling states above the condensate. In this respect, the operators $\hat{a}_0$ and  $\hat{a}^{+}_0$ are replaced by c-numbers  $\hat{a}_0=\hat{a}^{+}_0=\sqrt{N_0}$  within approximation the presence of a macroscopic number of condensate atoms $N_0\gg 1$. This assumption leads to a broken  of Bose-symmetry law for atoms occupying in the condensate state. To be refused a broken from Bose-symmetry law for bosons in the condensate, we apply the Penrose-Onsager's definition of the Bose condensation [10]:

\begin{equation}
\lim_{N_0, N\rightarrow\infty}\frac{N_0}{N}=const
\end{equation}

This reasoning is a very important factor for microscopic investigation of the model non-ideal Bose gas because the presence of a macroscopic number of atoms in the condensate means $N_0>>N_{\vec{p}\not =0}$  (where $ N_{\vec{p}\not =0}$ is the occupation number of  atoms with momentum  $\vec{p}\not =0$). In this respect, we may postulate a following approximation for an occupation number of  atoms with momentum  $\vec{p}$:    

\begin{equation}
\lim_{N_0\rightarrow\infty}\frac{N_{\vec{p}}}{N_0}=\delta_{\vec{p},0}
\end{equation}

The next step is to find the property of operators 
$\frac{\hat{a}^{+}_{\vec{p}_1-\vec{p}}}{\sqrt{N_0}}$, 
$\frac{\hat{a}_{\vec{p}_1-\vec{p}}}{\sqrt{N_0}}$ by applying (5). Obviously,

\begin{equation}
\lim_{N_0\rightarrow\infty}\frac{\hat{a}^{+}_{\vec{p}_1-\vec{p}}}{\sqrt{N_0}}=\delta_{\vec{p}_1, \vec{p}} 
\end{equation} 
and
\begin{equation}
\lim_{N_0\rightarrow\infty}\frac{\hat{a}_{\vec{p}_1-\vec{p}}}{\sqrt{N_0}}=
\delta_{\vec{p}_1, \vec{p}} 
\end{equation}

Excluding the term $\vec{p}_1=0$, :the density operators of bosons
$\hat{\varrho}_{\vec{p}}$ and $\hat{\varrho}^{+}_{\vec{p}}$ take the following forms:

\begin{equation}
\hat{\varrho}_{\vec{p}}= \sqrt{N_0}\biggl(\hat{a}^{+}_{-\vec{p}}+
\sqrt{2}\hat{c}_{\vec{p}}\biggl)
\end{equation}

and
\begin{equation}
\hat{\varrho}^{+}_{\vec{p}}=\sqrt{N_0}\biggl(\hat{a}_{-\vec{p}}+
\sqrt{2}\hat{c}^{+}_{\vec{p}}\biggl)
\end{equation}

where $\hat{c}_{\vec{p}}$ and $\hat{c}^{+}_{\vec{p}}$ are,
respectively, the Bose-operators of density-quasiparticles presented in reference [8] which in turn are the Bose-operators of bosons used in expressions (6) and (7):
:
\begin{equation}
\hat{c}_{\vec{p}}=\frac{1}{\sqrt{2N_0}} \sum_{\vec{p}_1\not=0}
\hat{a}^{+}_{\vec{p}_1-\vec{p}}\hat{a}_{\vec{p}_1}=\frac{1}{\sqrt{2}} \sum_{\vec{p}_1\not=0}\delta_{\vec{p}_1, \vec{p}} \hat{a}_{\vec{p}_1}=\frac{1}{\sqrt{2}}\hat{a}_{\vec{p}}
\end{equation}
and

\begin{equation}
\hat{c}^{+}_{\vec{p}} =\frac{1}{\sqrt{2N_0}}
\sum_{\vec{p}_1\not=0}\hat{a}^{+}_{\vec{p}_1}
\hat{a}_{\vec{p}_1-\vec{p}}=\frac{1}{\sqrt{2}} \sum_{\vec{p}_1\not=0}\delta_{\vec{p}_1, \vec{p}}\hat{a}^{+}_{\vec{p}_1}=
\frac{1}{\sqrt{2}}\hat{a}^{+}_{\vec{p}}
\end{equation}

Thus, we reach to the density operators of atoms $\hat{\varrho}_{\vec{p}}$ and 
$\hat{\varrho}^{+}_{\vec{p}}$, presented by Bogoliubov [3], at approximation $\frac{N_0}{N}=const$:

\begin{equation}
\hat{\varrho}_{\vec{p}}= \sqrt{N_0}\biggl(\hat{a}^{+}_{-\vec{p}}+
\hat{a}_{\vec{p}}\biggl)
\end{equation}

and
\begin{equation}
\hat{\varrho}^{+}_{\vec{p}}=\sqrt{N_0}\biggl(\hat{a}_{-\vec{p}}+
\hat{a}^{+}_{\vec{p}}\biggl)
\end{equation}

which displays that the density quasiparticles are absent.

The identical picture is observed in the case of the density excitations which was  predicted by the Glyde, Griffin and  Stirling  [5-7] where was proposed a presentation $\hat{\varrho}_{\vec{p}}$ in a following form: 

\begin{equation}
\hat{\varrho}_{\vec{p}}= \sqrt{N_0}\biggl(\hat{a}^{+}_{-\vec{p}}+
\hat{a}_{\vec{p}}+\tilde{\varrho}_{\vec{p}}\biggl)
\end{equation}
where terms involving $\vec{p}_1\not=0$ and $, \vec{p}_1\not=\vec{p}$  are 
written separately; and the operator $\tilde{\varrho}_{\vec{p}}$ describes  
the density-excitations: 

\begin{equation}
\tilde{\varrho}_{\vec{p}}=\frac{1}{\sqrt{N_0}} \sum_{\vec{p}_1\not=0, \vec{p}_1\not=\vec{p}}
\hat{a}^{+}_{\vec{p}_1-\vec{p}}\hat{a}_{\vec{p}_1}
\end{equation}

At inserting of (6) and (7) into (15), the term, representing as the density-excitations, vanishes because $\tilde{\varrho}_{\vec{p}}=0$.

Consequently, the Hamiltonian of system, presented in (1) by support of (12) and (13), reproduces an extension form of the Bogoliubov Hamiltonian, at approximation $\frac{N_0}{N}=const$:

\begin{equation}
\hat{H}_a = \sum_{\vec{p}\not=0}\biggl (\frac{p^2}{2m}+ mv^2\biggl) 
\hat{a}^{+}_{\vec{p}}\hat{a}_{\vec{p}}+
\frac{mv^2}{2}\sum_{\vec{p}\not=0} \biggl (\hat{a}^{+}_{-\vec{p}}\hat{a}^{+}_{\vec{p}}+\hat{a}_{\vec{p}}
\hat{a}_{-\vec{p}}\biggl)
\end{equation}

where $v=\sqrt{ \frac{ U_{\vec{p}} N_0}{mV}}=\sqrt{ \frac{4\pi d\hbar^2 N_0}{m^2V}}$ is the
velocity  of sound in the Bose gas which depends on the density atoms in the condensate $\frac{ N_0}{V}$.

For evolution of the energy level it is a necessary to diagonalize the Hamiltonian $\hat{H}_a$ which is accomplished by introduction of the Bose-operators $\hat{b}^{+}_{\vec{p}}$ and $\hat{b}_{\vec{p}}$ by using of the Bogoliubov linear transformation [3]:

\begin{equation} 
\hat{a}_{\vec{p}}=\frac{\hat{b}_{\vec{p}} + 
L_{\vec{p}}\hat{b}^{+}_{-\vec{p}}} {\sqrt{1-L^2_{\vec{p}}}} 
\end{equation}
where $L_{\vec{p}}$ is the unknown real symmetrical function
of  a momentum $\vec{p}$.

Substitution of (17) into (16) leads to 

\begin{equation}
\hat{H}_a=\sum_{\vec{p}}\varepsilon_{\vec{p}}\hat{b}^{+}_{\vec{p}}
\hat{b}_{\vec{p}}
\end{equation}

hence we infer that $\hat{b}^{+}_{\vec{p}}$ and
$\hat{b}_{\vec{p}}$ are the "creation" and "annihilation"
operators of a Bogoliubov quasiparticles with energy:

\begin{equation}
\varepsilon_{\vec{p}}=\biggl[\biggl(\frac{p^2}{2m}\biggl)^2 + p^2
v^2\biggl]^{1/2}
\end{equation}
 
In this context, the real symmetrical function $L_{\vec{p}}$
of  a momentum $\vec{p}$ is found

\begin{equation}
L^2_{\vec{p}}=\frac{\frac{p^2}{2m}+ mv^2-\varepsilon_{\vec{p}}}
{\frac{p^2}{2m}+
mv^2+\varepsilon_{\vec{p}}}
\end{equation}
 
It is well known, the strong interaction between the helium 
atoms is very important and reduces the condensate fraction 
to 10 percent or $\frac{N_0}{N}=0.1$  [4], at absolute zero. However, as we suggest a new model of dilute Bose gas, proposed herein, may have an significant application for describing of thermodynamic properties of the superfluid liquid helium because the S-wave scattering between two atoms, with coordinates $\vec{r}_1$ and $\vec{r}_2$ in the space of coordinate, is presented by the repulsive potential delta-function $U_{\vec{r}} =\frac{4\pi d \hbar^2\delta_{\vec{r}}}{m }$ from $\vec{r}=\vec{r}_1-\vec{r}_2$. On other hand, the presented model works on the condensed fraction $\frac{N_0}{N}\ll 1$ in differ from the Bogoliubov model where $\frac{N_0}{N}\approx 1$.

\vspace{5mm}
{\bf 3. The effective attractive potential interaction between neutron modes in a helium liquid-neutron gas mixture.}
\vspace{5mm}
 
We now attempt to describe the thermodynamic property of a helium liquid-neutron gas mixture.  In this context, we consider a neutron gas as an ideal Fermi gas consisting of  $n$ free neutrons with mass $m_n $ which interact with $N$ interacting atoms of a superfluid liquid helium. The helium-neutron mixture is confined in a box of volume $V$. The Hamiltonian of a considering system $\hat{H}_{a,n}$ consists of the term of the Hamiltonian of Bogoliubov excitations $\hat{H}_{a}$ in (18) and the term of the Hamiltonian of an ideal Fermi neutron gas as well as the term of interaction between the density of the Bogoliubov excitations and the density of the neutron modes: 
\begin{equation}
\hat{H}_{a,n}=\sum_{\vec{p},\sigma }\frac{p^2}{2m_n}
\hat{a}^{+}_{\vec{p},\sigma }\hat{a}_{\vec{p},\sigma} +\sum_{\vec{p}}\varepsilon_{\vec{p}}\hat{b}^{+}_{\vec{p}}
\hat{b}_{\vec{p}}+\frac{1}{2V}\sum_{\vec{p}\not=0}U_0 \hat{\varrho}_{\vec{p}}\hat{\varrho}_{-\vec{p},n }
\end{equation}

where  $\hat{a}^{+}_{\vec{p},\sigma}$ and 
$\hat{a}_{\vec{p},\sigma }$ are, respectively,  the operators of creation and 
annihilation for free neutron with momentum $\vec{p}$, by the value of its 
spin z-component $\sigma=^{+}_{-}\frac{1}{2}$; $U_0$ is the Fourier
transform of the repulsive interaction between  the density of the Bogoliubov excitations and the density modes of the neutrons: 
\begin{equation}
U_0 =\frac{4\pi d_0\hbar^2}{\mu} 
\end{equation}

where $d_0$ is the scattering amplitude between a helium atoms and neutrons; $\mu =\frac{m\cdot m_n}{m+m_n}$ is the relative mass.

Hence, we note that the Fermi operators $\hat{a}^{+}_{\vec{p},\sigma}$ and $\hat{a}_{\vec{p},\sigma }$ satisfy to 
the Fermi commutation relations $[\cdot\cdot\cdot]_{+}$ as:

\begin{equation}
\biggl[\hat{a}_{\vec{p},\sigma}, \hat{a}^{+}_{\vec{p}^{'},
\sigma^{'}}\biggl]_{+} =
\delta_{\vec{p},\vec{p^{'}}}\cdot\delta_{\sigma,\sigma^{'}}
\end{equation}

\begin{equation}
[\hat{a}_{\vec{p},\sigma}, \hat{a}_{\vec{p^{'}}, \sigma^{'}}]_{+}= 0
\end{equation}

\begin{equation}
[\hat{a}^{+}_{\vec{p},\sigma}, \hat{a}^{+}_{\vec{p^{'}}, 
\sigma^{'}}]_{+}= 0
\end{equation}

The density operator of neutrons with spin $\sigma$ in momentum 
$\vec{p}$ is defined as
\begin{equation}
\hat{\varrho}_{\vec{p},n }=\sum_{\vec{p}_1, \sigma }
\hat{a}^{+}_{\vec{p}_1-\vec{p},\sigma }\hat{a}_{\vec{p}_1,\sigma }
\end{equation}
where $\hat{\varrho}^{+}_{\vec{p},n }=\hat{\varrho}_{-\vec{p},n }$

The operator of total number of neutrons is
\begin{equation}
\sum_{\vec{p},\sigma}\hat{a}^{+}_{\vec{p},\sigma}\hat{a}_{\vec{p},\sigma}=
\hat{n}
\end{equation}
On other hand, the density operator, in the term of the Bogoliubov quasiparticles $\hat{\varrho}_{\vec{p}}$ included in (21), is expressed by following form, to application (17) into (12):

\begin{equation}
\hat{\varrho}_{\vec{p}}= \sqrt{N_0}\sqrt{\frac{1+ L_{\vec{p}}}{1- L_{\vec{p}}}}\biggl(\hat{b}^{+}_{-\vec{p}}+
\hat{b}_{\vec{p}}\biggl)
\end{equation}

Hence, we note that the Bose- operator  $\hat{b}_{\vec{p}}$ commutates with  the Fermi operator $\hat{a}_{\vec{p},\sigma }$ because the Bogoliubov excitations and neutrons are an independent.

Now, inserting of a value of operator $\hat{\varrho}_{\vec{p}}$ from (28) into (21), which in turn leads to reducing the Hamiltonian of system $\hat{H}_{a,n}$:

\begin{eqnarray}
\hat{H}_{a,n}&=&\sum_{\vec{p},\sigma }\frac{p^2}{2m_n}
\hat{a}^{+}_{\vec{p},\sigma }\hat{a}_{\vec{p},\sigma}+ \sum_{\vec{p}}\varepsilon_{\vec{p}}
\hat{b}^{+}_{\vec{p}}\hat{b}_{\vec{p}}+\nonumber\\
&+&\frac{ U_0\sqrt{N_0}}{2V}\sum_{\vec{p}} \sqrt{\frac{1+ L_{\vec{p}}}{1- L_{\vec{p}}}}\biggl(\hat{b}^{+}_{-\vec{p}}+\hat{b}_{\vec{p}}\biggl)\hat{\varrho}_{-\vec{p},n }
\end{eqnarray}

To allocate anomalous term in the Hamiltonian of system $\hat{H}_{a,n}$,  which denotes by third term in right side in (29), we apply a canonical transformation for
the operator $\hat{H}_{a,n}$ within introducing a new operator $\tilde{H}$:

\begin{equation}
\tilde{H}=\exp\biggl(\hat{S}^{+}\biggl)\hat{H}_{a,n}
\exp\biggl(\hat{S}\biggl)
\end{equation}

which is decayed by following terms: 

\begin{equation}
\tilde{H}=\exp\biggl(\hat{S}^{+}\biggl)\hat{H}_{a,n}
\exp\biggl(\hat{S}\biggl) = \hat{H}_{a,n}-[\hat{S},\hat{H}_{a,n}]+
\frac{1}{2}[\hat{S},[\hat{S},\hat{H}_{a,n}]]-\cdots
\end{equation}

where the operators represent as:
\begin{equation}
\hat{S}^{+}=\sum_{\vec{p}}\hat{S^{+}_{\vec{p}}}
\end{equation}
and   
\begin{equation}
\hat{S}=\sum_{\vec{p}}\hat{S_{\vec{p}}}
\end{equation}
and satisfy  to a condition $\hat{S}^{+} = -\hat{S}$

In this respect, we assume that

\begin{equation}
\hat{S_{\vec{p}}}=A_{\vec{p}}\biggl (\hat{\varrho}_{\vec{p},n}
\hat{b}_{\vec{p}}-
\hat{\varrho}^{+}_{\vec{p},n }\hat{b}^{+}_{\vec{p}}\biggl)
\end{equation}

where $A_{\vec{p}}$ is the unknown  real symmetrical function from  a momentum
$\vec{p}$. In this context, at application $\hat{S_{\vec{p}}}$ from (34) to (33) with taking into account $\hat{\varrho}^{+}_{-\vec{p},n }=\hat{\varrho}_{\vec{p},n }$, then we obtain 
\begin{equation}
\hat{S}=\sum_{\vec{p}}\hat{S_{\vec{p}}}= 
\sum_{\vec{p}}A_{\vec{p}}\hat{\varrho}_{\vec{p},n}\biggl (
\hat{b}_{=\vec{p}}-\hat{b}^{+}_{\vec{p}}\biggl)
\end{equation}
In analogy manner, at $\hat{\varrho}^{+}_{-\vec{p},n }=\hat{\varrho}_{\vec{p},n }$, we have
\begin{equation}
\hat{S}^{+}=\sum_{\vec{p}}\hat{S^{+}_{\vec{p}}}=
\sum_{\vec{p}}A_{\vec{p}}\hat{\varrho}^{+}_{\vec{p},n}\biggl (
\hat{b}^{+}_{\vec{p}}-\hat{b}_{-\vec{p}}\biggl)=
- \sum_{\vec{p}}A_{\vec{p}}\hat{\varrho}_{\vec{p},n}\biggl (
\hat{b}_{-\vec{p}}-\hat{b}^{+}_{\vec{p}}\biggl)
\end{equation}

To find  $A_{\vec{p}}$, we substitute  (29), (35) and (36) into  (31). Then,

\begin{equation}
[\hat{S},\hat{H_{a,n} }]=\frac{1}{V}\sum_{\vec{p}} 
A_{\vec{p}}U_0\sqrt{N_0}\sqrt{\frac{1+ L_{\vec{p}}}
{1- L_{\vec{p}}}}\hat{\varrho}_{\vec{p},n}
\hat{\varrho}_{-\vec{p},n}+
\sum_{\vec{p}} A_{\vec{p}} 
\varepsilon_{\vec{p}}\biggl(\hat{b}^{+}_{\vec{p}}+\hat{b}_{-\vec{p}}\biggl) \hat{\varrho}_{-\vec{p},n}
\end{equation}

\begin{equation}
\frac{1}{2}[\hat{S},[\hat{S},\hat{H}_{a,n}]]=
\sum_{\vec{p}} A^2_{\vec{p}} 
\varepsilon_{\vec{p}}\varrho_{\vec{p},n}
\hat{\varrho}_{-\vec{p},n}
\end{equation}

and $[\hat{S}, [\hat{S},[\hat{S},\hat{H}_{a,n}]]]=0$ within application a Bose commutation relations as 
$[\varrho_{\vec{p}_1,n},\hat{\varrho}_{\vec{p}_2,n}]=0$ and $[\hat{a}^{+}_{\vec{p}_1,\sigma}\hat{a}_{\vec{p}_1,\sigma},
\hat{\varrho}_{\vec{p}_2,n}]=0$. 

Thus, the form of new operator $\tilde{H}$ in (31) takes a following form:

\begin{eqnarray}
\tilde{H}& =&\sum_{\vec{p}}\varepsilon_{\vec{p}}
\hat{b}^{+}_{\vec{p}}\hat{b}_{\vec{p}}+
\frac{1}{2V}\sum_{\vec{p}}U_0\sqrt{N_0}
\sqrt{\frac{1+ L_{\vec{p}}}{1- L_{\vec{p}}}}
\biggl(\hat{b}^{+}_{-\vec{p}}+\hat{b}_{\vec{p}}\biggl)
\hat{\varrho}_{-\vec{p},n }+
\nonumber\\
&+&\sum_{\vec{p},\sigma }\frac{p^2}{2m_n}
\hat{a}^{+}_{\vec{p},\sigma }
\hat{a}_{\vec{p},\sigma}-
\frac{1}{V}\sum_{\vec{p}} A_{\vec{p}}U_0
\sqrt{N_0}\sqrt{\frac{1+ L_{\vec{p}}}
{1- L_{\vec{p}}}}\hat{\varrho}_{\vec{p},n }
\hat{\varrho}_{-\vec{p},n}-
\nonumber\\
&-&\sum_{\vec{p}} A_{\vec{p}} 
\varepsilon_{\vec{p}}
\biggl(\hat{b}^{+}_{-\vec{p}}+\hat{b}_{\vec{p}}\biggl) 
\hat{\varrho}_{-\vec{p},n }
+\sum_{\vec{p}} A^2_{\vec{p}} 
\varepsilon_{\vec{p}}\hat{\varrho}_{\vec{p},n }
\hat{\varrho}_{-\vec{p},n}
\end{eqnarray}

The transformation of the term of the interaction between the density of the Bogoliubov modes and the density neutron modes is made by removing of a second and fifth terms in right side of (39) which leads to obtaining of a quantity for
$A_{\vec{p}}$:

\begin{equation}
A_{\vec{p}}=\frac{U_0 \sqrt{N_0}}{2\varepsilon_{\vec{p}} V}\cdot\sqrt{\frac{1+ L_{\vec{p}}}{1- L_{\vec{p}}}}
\end{equation}

In this respect, we reach to reducing of the new Hamiltonian of system (39):

\begin{eqnarray}
\tilde{H}& =&\sum_{\vec{p}}\varepsilon_{\vec{p}}
\hat{b}^{+}_{\vec{p}}\hat{b}_{\vec{p}}+
\sum_{\vec{p},\sigma }\frac{p^2}{2m_n}
\hat{a}^{+}_{\vec{p},\sigma }\hat{a}_{\vec{p},\sigma}-\nonumber\\
&-&\frac{1}{V}\sum_{\vec{p}} A_{\vec{p}}U_0\sqrt{N_0}\sqrt{\frac{1+ L_{\vec{p}}}{1- L_{\vec{p}}}}\hat{\varrho}_{\vec{p},n }\hat{\varrho}_{-\vec{p},n}+\sum_{\vec{p}} A^2_{\vec{p}} \varepsilon_{\vec{p}}\hat{\varrho}_{\vec{p},n }\hat{\varrho}_{-\vec{p},n}
\end{eqnarray}

As result, the new form of Hamiltonian system takes a following fom:

\begin{equation}
\tilde{H}= \sum_{\vec{p}}\varepsilon_{\vec{p}}
\hat{b}^{+}_{\vec{p}}\hat{b}_{\vec{p}}+\hat{H}_n
\end{equation}
where $\hat{H}_n $ is the effective Hamiltonian of 
a neutron gas which contains an effective interaction between neutron modes:

\begin{equation}
\hat{H}_n =\sum_{\vec{p},\sigma }\frac{p^2}{2m_n}
\hat{a}^{+}_{\vec{p},\sigma }\hat{a}_{\vec{p},\sigma} +\frac{1}{2V}\sum_{\vec{p}}V_{\vec{p}}
\hat{\varrho}_{\vec{p},n}\hat{\varrho}_{-\vec{p},n}
\end{equation}

where $V_{\vec{p}}$ is the effective potential of the interaction between neutron modes which takes a following form at substituting a value of $ A_{\vec{p}}$ from (40) into (41): 

\begin{equation}
V_{\vec{p}}= -2A_{\vec{p}}U_0\sqrt{N_0}\sqrt{\frac{1+ L_{\vec{p}}}{1- L_{\vec{p}}}}+ 2A^2_{\vec{p}} \varepsilon_{\vec{p}}V=-\frac{ U^2_0 N_0\biggl (1+ L_{\vec{p}}\biggl )}{ V \varepsilon_{\vec{p}} \biggl (1- L_{\vec{p}}\biggl )}
\end{equation}

In this letter, we consider following cases:
1. At low momenta atoms of a helium  $p<<2mv$, the Bogoliunov's quasiparticles in (19) represent as the phonons with energy $\varepsilon_{\vec{p}}\approx pv$  which in turn defines a value $L^2_{\vec{p}}\approx \frac{1-\frac{p}{mv}}{1+\frac{p}{mv}}\approx \biggl(1-\frac{p}{mv}\biggl)^2$ in (20)  or  $L_{\vec{p}}\approx 1-\frac{p}{mv}$. In this context, the effective potential between neutron modes takes a following form:

\begin{equation}
V_{\vec{p}}\approx -\frac{2 m U^2_0 N_0}{V p^2}= -\frac{4\pi \hbar^2 e^2_1}{p^2}
\end{equation}

The value $ e_1$ is the effective charge, at a small momenta of atoms: 

$$
e_1=\frac{U_0}{\hbar }\sqrt{\frac{ m N_0}{2V \pi }}
$$

2. At high momenta atoms of a helium $p>>2mv$, we obtain $\varepsilon_{\vec{p}}\approx \frac{p^2}{2m}+ mv^2$ in (19) which in turn defines $L_{\vec{p}}\approx 0$ in (20).  Then, the effective potential between neutron modes presents as:

\begin{equation}
V_{\vec{p}}\approx -\frac{ m U^2_0 N_0}{V p^2}= -\frac{4\pi \hbar^2 e^2_2}{p^2}
\end{equation}

where $e_2$ is the effective charge, at high momenta of atoms:
$$
e_2=\frac{U_0}{2\hbar }\sqrt{\frac{ m N_0}{V \pi }}
$$
  
Consequently, in both cases, the effective scattering between two neutrons is presented in the coordinate space by a following form:

\begin{equation}
V (\vec{r})=\frac{1}{V}\sum_{\vec{p}} V_{\vec{p}}\cdot e^{i\frac{\vec{p}\vec{r}}{\hbar}}=-\frac{e^2_*}{r}
\end{equation}
where $ e_*= e_1$, at small momenta of atoms; and $ e_*= e_2$, at high momenta. 

\vspace{5mm}
{\bf 4. Creation Spinless Neutron Pairs.}
\vspace{5mm}

The term of the interaction between two neutrons $V (\vec{r})$ in the coordinate space mediates the attractive Coulomb interaction between two charged particles with mass of neutron $m_n$, having the opposite effective charges $ e_*$ and $- e_*$, which together create a neutral system. Indeed, the effective Hamiltonian of a neutron gas in (43) is rewrite down in the space of coordinate by following form:

\begin{equation}
\hat{H}_n =\sum^{\frac{n}{2}}_{i=1}\hat{H}_i =-\frac{\hbar^2}{2m_n}\sum^{n}_{i=1} \Delta_i-\sum_{i<j}\frac{e^2_*}{\mid\vec{r}_i-\vec{r}_j\mid }
\end{equation}

where $\hat{H}_i $ is the Hamiltonian of system consisting two neutron with opposite spin which have a coordinates $\vec{r}_i $ and $\vec{r}_j $:
\begin{equation}
\hat{H}_i=-\frac{\hbar^2}{2m_n}\Delta_i-\frac{\hbar^2}{2m_n}\Delta_j-\frac{e^2_*}{\mid\vec{r}_i-\vec{r}_j\mid }
\end{equation}
The transformation of considering coordinate system to the relative coordinate $\vec{r}=\vec{r}_i-\vec{r}_j $ and the coordinate of center mass $\vec{R}=\frac{\vec{r}_i+\vec{r}_j }{2}$, we have 

\begin{equation}
\hat{H}_i=-\frac{\hbar^2}{4m_n}\Delta_R-\frac{\hbar^2}{m_n}\Delta_r-\frac{e^2_*}{r}
\end{equation}

In analogy of the problem Hydrogen atom, two neutrons with opposite spins is bound as a spinless neutron pair with mass 
$m_0=2m_n$ and with binding energy:
 
\begin{equation}
E_n=-\frac{m_n e^4_*}{4\hbar^2 n^2}=-\frac{const}{ n^2}\cdot \biggl(\frac{N_0}{V}\biggl)^2
\end{equation}

where $n$ is the main quantum number which determines a bound state on a neutron pair, at $ const >0$.

Thus, the spinless neutron pair is created in a helium liquid-neutron gas mixture by the term of the interaction between the Bogoliubov excitations and neutron modes which is removed by an induced the effective interaction which mediate via neutron modes. The later determines a bound state on 
a neutron pair with binding energy (51) which depends on the density of atoms in the condensate $\frac{N_0}{V}$, and therefore, may define the state of temperatures $0\leq T < T_c$ (where $T_c$ is the critical temperature of the Bose gas) for existing of neutron pairs. In accordance with this reasoning, the new Hamiltonian system takes a following form:

\begin{equation}
\tilde{H}= \sum_{\vec{p}}\varepsilon_{\vec{p}}
\hat{b}^{+}_{\vec{p}}\hat{b}_{\vec{p}}+\sum_{\vec{p},\sigma }\frac{p^2}{2m_0}
\hat{d}^{+}_{\vec{p}}\hat{d}_{\vec{p}} 
\end{equation}
where $\hat{d}^{+}_{\vec{p}}$ and $\hat{d}_{\vec{p}}$ are, respectively, the
"creation" and  "annihilation" Bose-operators of a free neutron pair with 
momentum $\vec{p}$.

As we see two independent types spinless describe a helium-neutron mixture, which are represent as the Bogoliubov quasiparticles and the neutron pair modes
in the state of temperatures $0\leq T < T_c$. At temperatures $T\geq T_c$, the neutron pair is decayed on two free neutrons because the fraction of the condensate atoms takes a zero meaning $\frac{N_0}{V}=0$, and in turn the binding energy $ E_n =0$ in (51). In this respect, the Bogoliubov excitations represent as free atoms of helium in addition to a free neutron gas with the Hamiltonian of system: 

\begin{equation}
\tilde{H}= \sum_{\vec{p}}\frac{p^2}{2m}
\hat{a}^{+}_{\vec{p}}\hat{a}_{\vec{p}}+\sum_{\vec{p},\sigma }\frac{p^2}{2m_n}
\hat{a}^{+}_{\vec{p},\sigma }\hat{a}_{\vec{p},\sigma} 
\end{equation}

In conclusion, we note that the new model of Bose gas, 
presented in this letter, might be useful for describing 
of the thermodynamic properties of a dilute gas of the Boson-Fermion mixtures confined in traps. The presented model of a helium liquid-neutron gas mixture gives explanation of the presence of a broad component in inelastic neutron scattering intensity, at lambda transition [4] because there is an appearance transformation of   neutron pair modes, in addition to single neutron modes, which in turn determine a superfluid phase for helium.  
\newpage 
\begin{center} 
{\bf References} 
\end{center} 
 
\begin{enumerate} 

\item
F.~London~, Nature, ~{\bf 141},~643~(1938)
\item 
L.~Landau~, J. Phys.(USSR), ~{\bf 5},~77~(1941); Phys.(USSR),
~{\bf 11},~91~(1947).
\item 
N.N.~Bogoliubov~, Jour. of Phys.(USSR), ~{\bf 11},~23~(1947)
\item 
N.M.~Blagoveshchenskii~ et al.,Phys. Rev. B ~{\bf 50}, ~16550~(1994)
\item 
H.R.~Glyde~ and A.~Griffin~., Phys.Rev.Lett.~{\bf 65},~1454~(1990). 
\item 
W.G.~Stirling~,H.R.~Glyde~,  Phys.Rev.B. ~{\bf 41},~4224~(1990) 
\item 
H.R.~Glyde~, Phys.Rev.B. ~{\bf 45},~7321~(1992)
\item 
V.N.~Minasyan~ et.al~,~ Phys.Rev.Lett.~{\bf 90},~235301~(2003)
\item
K. ~Morawetz ~et al,, ~Phys. Rev. B~ 76~, ~075116~ (2007)
\item 
O. ~Penrose~ and L. ~Onsager~, ~Phys. Rev.,~{\bf 104}~, ~576~ (1956)
\end{enumerate} 
\end{document}